\documentclass[a4paper,10pt]{article}
\usepackage{enumerate}
\usepackage{color}
\usepackage[utf8]{inputenc} 
\usepackage[english]{babel}
\usepackage[T1]{fontenc}
\usepackage{graphicx}
\usepackage{amsfonts,amssymb,amsmath,latexsym,amsthm}
\usepackage{textcomp}
\usepackage[pdftex]{hyperref}
\usepackage{geometry}
\DeclareMathOperator{\sech}{sech}

\title{Non-integrable soliton gas: The Schamel equation framework}
\author{Marcelo V. Flamarion$^{1}$, Efim Pelinovsky$^{2,3}$ and Ekaterina Didenkulova$^{2,3}$}
\date{}

\begin{document}
\maketitle
\begin{center}
{\footnotesize $^1$Unidade Acad{\^ e}mica do Cabo de Santo Agostinho, \\
UFRPE/Rural Federal University of Pernambuco, BR 101 Sul, Cabo de Santo Agostinho-PE, Brazil,  54503-900 \\
marcelo.flamarion@ufrpe.br }

\vspace{0.3cm}
{\footnotesize 
$^{2}$Faculty of Informatics, Mathematics and Computer Science, HSE University, Nizhny Novgorod 603155, Russia.

$^{3}$ Il'ichev Pacific Oceanological Institute Far Eastern Branch Russian Academy of Sciences, Vladivostok 690041, Russia.}



\end{center}


\begin{abstract} 
Soliton gas or soliton turbulence is a subject of intense studies due to its great importance to optics, hydrodynamics, electricity, chemistry, biology and plasma physics. Usually,  this term is used for integrable models where solitons interact elastically. However, soliton turbulence can also be a part of non-integrable dynamics, where long-lasting solutions in the form of almost solitons may exist. In the present paper, the complex dynamics of ensembles of solitary waves is studied within the Schamel equation using direct numerical simulations. Some important statistical characteristics  (distribution functions, moments) are calculated numerically for unipolar and bipolar soliton gases. Comparison of results with integrable Korteweg-de Vries (KdV) and modified KdV (mKdV)  models are given qualitatively. Our results agree well with the predictions of the KdV equation in the case of unipolar solitons. However, in the bipolar case, we observed a notable departure from the mKdV model, particularly in the behavior of kurtosis. The observed increase in kurtosis signifies the amplification of distribution function tails, which, in turn, corresponds to the presence of high-amplitude waves.

	\end{abstract}

\section{Introduction}
Solitons are the exact solutions of many equations and have many applications in nonlinear dynamics, including optical fibers, surface and internal waves in the ocean, laboratory and astrophysical plasma, etc. According to the classic definition, they are coherent large amplitude pulses whose shape and speed are not changed their interactions. Such particle-like behavior is explained by the balance between dispersion and nonlinearity, which from one hand tend to wave spreading and from the other – lead to its steepening. The great importance of solitons lies in their ability to transfer energy over long distances. Thus, the physical system where the solitons play the key role in the dynamics may be subjected to the formation of abnormally large waves (i.e. rogue waves or freak waves). This problem initially arose within integrable models such as the nonlinear Schrodinger and the KdV equations. V. Zakharov \cite{Zakharov:1971, Zakharov:2009}  first introduced the concept of soliton turbulence  where the kinetic theory of rarefied solitons was build. Later on, this concept has been extended to the dense soliton gas with frequently interacted solitons \cite{G:2005, G:2021}. Kinetic equations describes the transport of spectral data of the associated scattering problem, but there is no information about phases (polarity) of solitons and wave fields themselves, therefore, they are unsuitable for the study of their statistics which has practical importance. Direct numerical simulation of wave ensemble became a convenient alternative. These results for integrable KdV-models can be found in \cite{Shurgalina:2016a, Shurgalina:2017, Shurgalina:2016b, Pelinovsky:2016, Didenkulova:2019, Slunyaev:2022a, Slunyaev:2022b}. Similar studied concerned another type of wave, preserving their energy – wave packets called breathers. Their collective dynamics promotes freak wave formation \cite{Didenkulova:2022, Slunyaev:2018}. Soliton and breather turbulence is extensively investigated within the nonlinear Schrodinger equation in the contexts of water wave dynamics and nonlinear optics \cite{Gelash:2018, Crespo:2016, Zakharov:2015, Randoux:2016, Copie:2020, Walczak:2015, El:2020}. Moreover, the presence of soliton and breather turbulence in ocean waves was confirmed in \cite{Costa:2014, Osborne:2019}.

The problem of soliton turbulence may also be investigated in non-integrable models, allowing the existence of soliton-like impulses, which interact almost elastically. Dutykh and Pelinovsky \cite{Dutykh:2014}  compared the collective behaviour of soliton ensembles within the KdV equation the non-integrable KdV–BBM type models using  direct numerical simulations. The closeness in the behavior of the wave fields was ascertained, including the fact that the probability distributions remain quasi-invariant during the system evolution for both KdV and KdV-BBM cases. In the present paper we  study of soliton turbulence to the Schamel equation, which is not integrable by the inverse scattering transform since a Lax pair does not exist for this equation. It describes the development of a localized, coherent wave structure that propagates in plasma \cite{Schamel:1973, Schamel:2017, Williams:2014, Cheemaa:2019}. This equation contains modular nonlinear term with non-integer power, and this stands out strongly on background of traditional equations Korteweg-de Vries hierarchy. The features of two soliton collision in the framework of the Schamel equation were resently investigated in \cite{Flamarion:2023}. It was shown that the soliton interaction follows the classical scenario of the modified Korteweg-de Vries soliton with small difference due to non-integrability of the Shamel equation. In the present work we study  the dynamics of the soliton gas within the Schamel equation and its statistical properties. In the section 2 the Schamel equation and  the numerical methods. In the Section 3 the collective dynamics of ensembles of solitons with the same polarity is considered. Further, in Section 4 we discover the features or bipolar soliton collision and statistical properties of bipolar soliton gas, with special emphasis on freak wave formation in such wave fields. Conclusion is given at the end of the paper.


\section{The Schamel equation}
In our research, we investigate solitary wave interactions by focusing on the Schamel equation in its canonical form
\begin{equation}\label{Schamel1}
u_{t} +\sqrt{|u|}u_{x}+u_{xxx}=0.
\end{equation}
Within this equation, the variable $u$ represents the wave field at a specific position $x$ and time $t$. It is worth noting that the Schamel equation is a Hamiltonian equation, meaning it possesses a Hamiltonian function that governs its behavior. The Hamiltonian associated with this equation is defined as follows
\begin{equation}\label{Hamiltonian}
\mathcal{H} = \int_{-\infty}^{+\infty}\Big[-\frac{1}{2}u_{x}^{2}+\frac{4}{15}\mathrm{sign}(u)|u|^{5/2}\Big] dx.
\end{equation}
By expressing Equation (\ref{Schamel1}) in Hamiltonian form with respect to the functional $\mathcal{H}$, we can establish a relationship between the wave dynamics and the Hamiltonian. This connection is expressed by the following equation
\begin{equation*}
u_{t} = \frac{\partial}{\partial x}\Bigg[\frac{\delta\mathcal{H}}{\delta u}\Bigg],
\end{equation*}
where the functional derivative of the Hamiltonian with respect to $u$ is given by
\begin{equation*}
\frac{\delta\mathcal{H}}{\delta u}=u_{xx}+\frac{2}{3}\mathrm{sign}(u)|u|^{3/2}.
\end{equation*}

One intriguing feature of the Schamel equation is the invariance of its Hamiltonian $\mathcal{H}$ due to the absence of explicit time dependence. This invariance implies that the Hamiltonian remains constant throughout the evolution of the wave system. Furthermore, the Schamel equation (\ref{Schamel1}) possesses an additional invariant, known as the Casimir invariant or the mass invariant. This quantity is defined by the following integral
\begin{equation}\label{mass}
M(t) = \int_{-\infty}^{+\infty}u(x,t) dx,
\end{equation}
and it characterizes the mass or the total "amount" of the wave field at any given time $t$. In addition to the mass invariant, the equation also exhibits a momentum invariant, given by
\begin{equation}\label{momentum}
P(t) = \int_{-\infty}^{+\infty}u^{2}(x,t) dx.
\end{equation}
These invariants, namely the Hamiltonian (\ref{Hamiltonian}), the mass (\ref{mass}), and the momentum (\ref{momentum}), play a crucial role in evaluating the accuracy and reliability of numerical methods employed to solve the Schamel equation (\ref{Schamel1}).

The Schamel equation (\ref{Schamel1}) supports solitary waves as its solutions. These solitary waves can be described by the following expressions
\begin{equation}\label{solitary}
u(x,t) = a\sech^{4}\left(k(x-ct)\right), \mbox{ where }  c = \frac{8\sqrt{|a|}}{15} \mbox{ and } k = \sqrt{\frac{c}{16}}.
\end{equation}
Here, $a$ stands for the amplitude of the solitary wave, which can be positive or negative. The parameter $c$ denotes the speed of the solitary wave and $k$ characterizes its wavenumber.

The KdV equation 
\begin{equation}\label{KdV}
u_{t} +uu_{x}+u_{xxx}=0.
\end{equation}
also admits solitary wave as solutions described by the formulas
\begin{equation}\label{solitaryKdV}
u(x,t) = a\sech^{2}\left(k(x-ct)\right), \mbox{ where }  c = \frac{a}{3} \mbox{ and } k = \sqrt{\frac{a}{12}}.
\end{equation}

\begin{figure}[h!]\
	\centering	
	\includegraphics[scale =0.99]{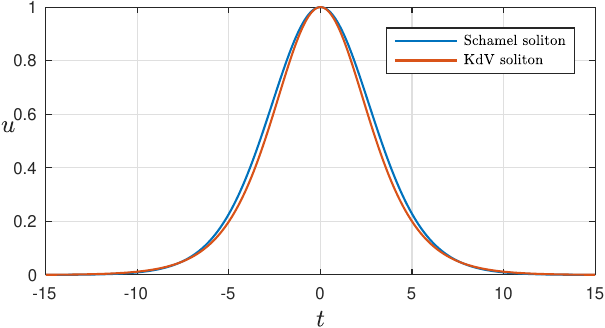}
	\caption{ Soliton profile of the Schamel equation (\ref{solitary})  and the KdV equation (\ref{solitaryKdV}).}
	\label{Solitons}
\end{figure}

Soliton solutions of the Schamel equation are wider than the KdV solitons (see Figure \ref{Solitons}), and they propagate faster than KdV ones.

The Schamel equation (\ref{Schamel1}) is solved numerically through a Fourier pseudospectral method combined with an integrating factor. The computational domain chosen for the simulation is a periodic interval $[-L, L]$, discretized with an equidistant grid consisting of $N$ points. This grid configuration facilitates precise approximation of spatial derivatives, as discussed in \cite{Trefethen:2001}. To mitigate the influence of spatial periodicity, a sufficiently large computational domain is employed. For the temporal evolution of the equation, the classical fourth-order Runge-Kutta method is employed with discrete time steps of size $\Delta t$. Typical simulations employ parameter values such as $L=200$, $N=2^{13}$, and $\Delta t =0.005$.  Numerical simulations are controlled by retaining of the first and second moments with precision of $10^{-9}$ and $10^{-8}$, respectively.

\section{Unipolar soliton}
\begin{figure}[h!]
\centering	
\includegraphics[scale=1]{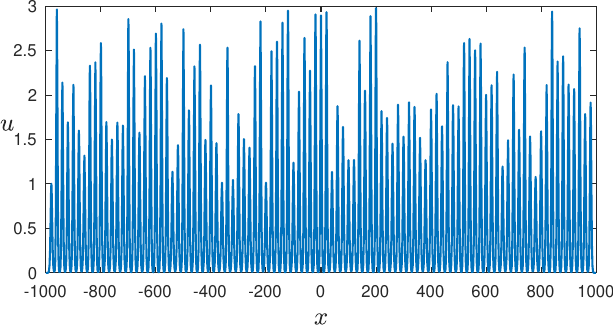}
\caption{ Initial distribution of  solitons.} 
\label{InitialDistribution}
\end{figure}

\begin{figure}[h!]
\centering	
\includegraphics[scale =1]{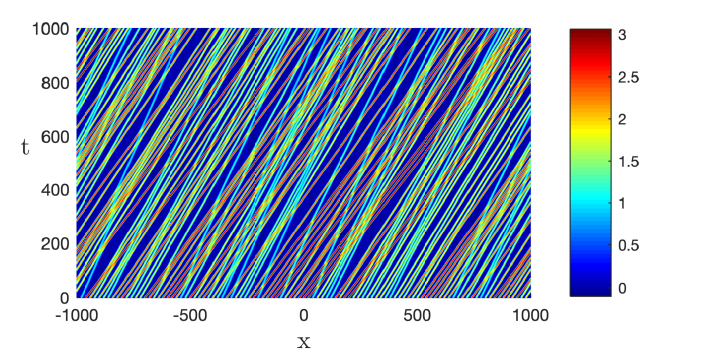}
\caption{ $x-t$ diagram of soliton field.} 
\label{mesh1}
\end{figure}

With aim to study the dynamics of unipolar soliton ensemble, we set initial wave field as a sequence of 100 separated solitons with amplitudes uniformly distributed from the range $[1, 3]$ in random order and fixed distance between their positions is $20$ units (see Figure \ref{InitialDistribution}). The solitons propagate to the right and owing to different speeds interact with each other. Non-dimensional time of calculation is set equal to 1000, so that solitons have time to interact. Multiple soliton interactions during the computational time is clearly seen in spatio-temporal diagram (Figure \ref{mesh1}). Due to the repulsion of unipolar solitons, pair soliton interactions predominate here, which were studied in detail in \cite{Flamarion:2023}. In the process of wave collision solitons get phase shifts, thus after several interactions their locations are hardly predicted

\begin{figure}[h!]
\centering	
\includegraphics[scale =1]{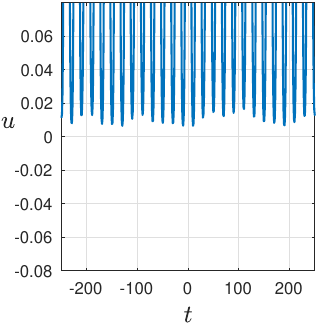}
\includegraphics[scale =1]{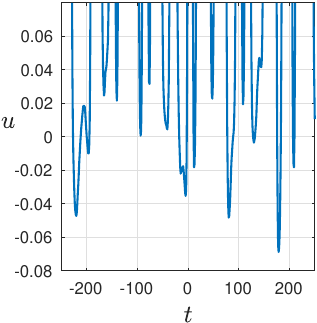}
\caption{ Zoom on portion of the computational domain at $t=0$ (left) and at $t=1000$ (right).} 
\label{Portion}
\end{figure}

\begin{figure}[h!]
\centering	
\includegraphics[scale =1]{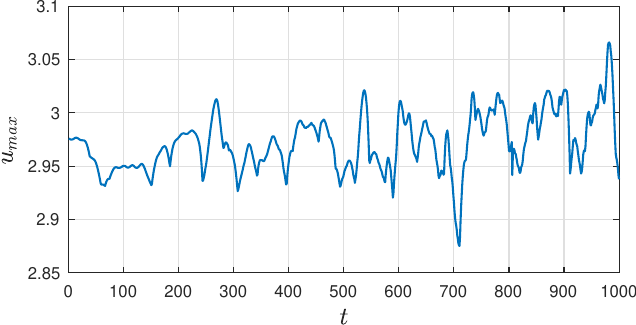}
\caption{ Temporal variability of the maxima of the wave field.} 
\label{TemporalUni}
\end{figure}

It is well known that interaction of unipolar KdV-like solitons leads to decrease in amplitude of resulting impulse \cite{Flamarion:2023, Pelinovsky:2013, Shurgalina:2018a,  Shurgalina:2018b, Shurgalina:2015, Anco:2015}. In the non-integrable Schamel equation this property is the same in case of two-soliton collision. However, there is a radiation created by inelastic collisions of solitons which is displayed in details in Figure \ref{Portion}. While the dispersive tail amplitudes are approximately only $1\%$ of the averaged amplitude of the initial soliton distribution, they still exert a minor influence, leading to a slight increase in the maximum wave field following interactions with other solitons (see Figure \ref{TemporalUni}). Here, the amplitude of the biggest initial soliton is 3, and in the Figure maximum of the wave field reaches 3.07. This distinguishes the unipolar gas of the Schamel equation from the unipolar gas of the integrable KdV equation. 

\begin{figure}[h!]
\centering	
\includegraphics[scale =1]{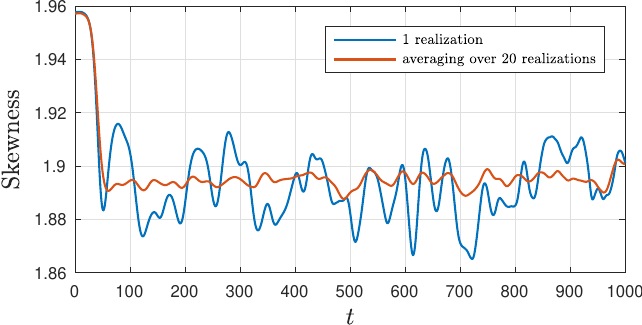}
\includegraphics[scale =1]{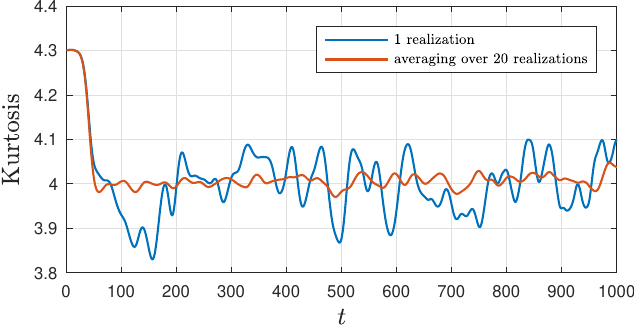}
\caption{Temporal evolution of the skewness and kurtosis of the unipolar soliton.} 
\label{MomentsUni}
\end{figure}

\begin{figure}[h!]
\centering	
\includegraphics[scale =1]{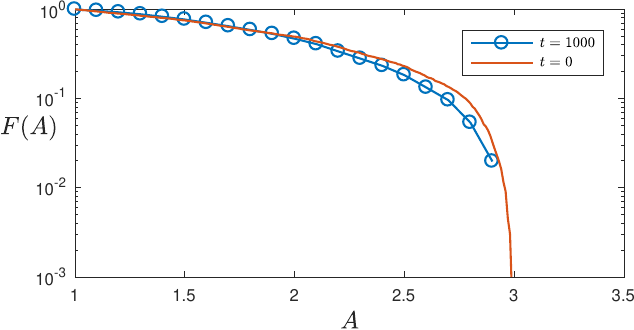}
\caption{Distribution function of wave amplitudes at different times averaged over 20 realizations.} 
\label{DistributionUni}
\end{figure}

\begin{figure}[h!]
\centering	
\includegraphics[scale =1]{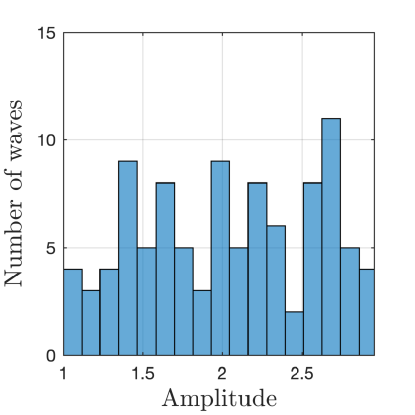}
\includegraphics[scale =1]{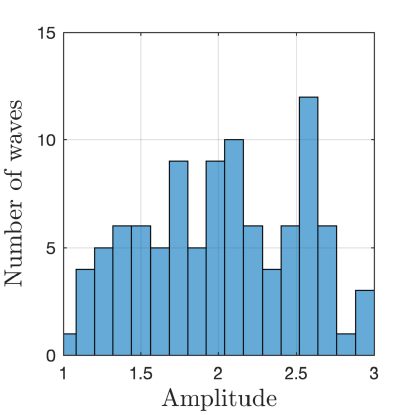}
\caption{Histogram of wave amplitude at $t=0$ (left) and at $t=1000$ (right).} 
\label{HistoUni}
\end{figure}

The fluctuations of the wave fields, which form after soliton collision, are small enough in order to influence higher statistical moments: skewness and kurtosis. Figure \ref{MomentsUni} demonstrates the temporal evolution of skewness and kurtosis of the unipolar soliton field for one realization and averaged value over 20 realizations. Similar to KdV model there is a short transition zone of sharp decrease of moments till about $t=50$ and it is the same for all realizations of soliton gas. Averaging over realization predictably diminishes the fluctuations of moments and it tends to stationary state.

The interactions among solitons have a discernible impact on the distribution functions of the wave field, as depicted in Figure \ref{DistributionUni}. Specifically, the average distribution function of wave amplitudes (corresponding to local maxima of the wave field) undergoes a downward shift in the high-amplitude region. Consequently, the presence of large waves diminishes, leading to a more uniform wave field. These findings align with similar observations made in studies involving soliton gases governed by the KdV and mKdV equations \cite{Dutykh:2014, Shurgalina:2016a}. The evolution of the number crests count for a single realization at different times is illustrated in Figure \ref{HistoUni}.



\section{Bipolar soliton}
\begin{figure}[h!]
\centering	
\includegraphics[scale =1]{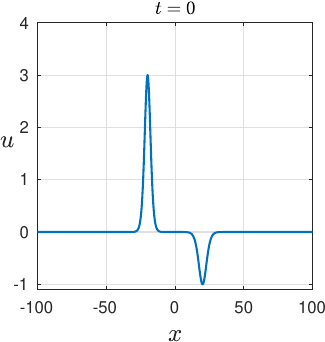}
\includegraphics[scale =1]{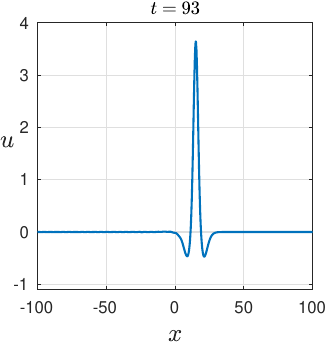}
\includegraphics[scale =1]{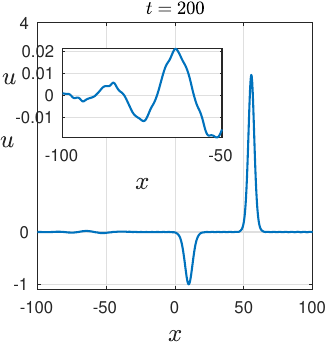}
\caption{Interaction process of bipolar soliton interaction within the Schamel equation.} 
\label{Collision}
\end{figure}

Presence of solitons with different polarity makes the dynamics of the wave system more extreme, because the interaction bipolar solitons increases the maximum of the wave field unlike interaction of unipolar solitons. Formation of abnormally large waves as a result of bipolar soliton collision in the mKdV equation was demonstrated in \cite{Shurgalina:2016a, Pelinovsky:2016, Slunyaev:2022b}. However, because of the presence of radiation the maximum of resulting impulse is less than superposition of amplitudes of interacted solitons in non-integrable systems. The process of bipolar soliton collision within the Schamel equation is presented in Figure \ref{Collision}.

\begin{figure}[h!]
\centering	
\includegraphics[scale =1]{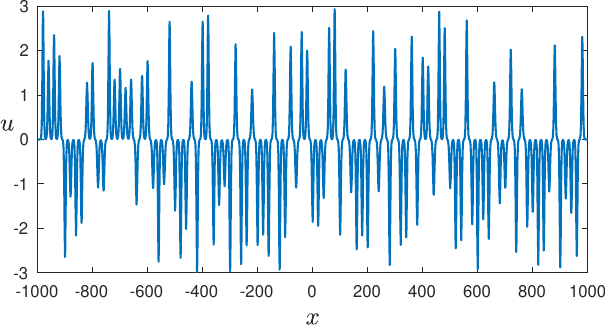}
\caption{ Initial distribution of bipolar  solitons.} 
\label{Bipolardist}
\end{figure}

To study the complex soliton dynamics with different polarities, we set initial wave field as a superposition of separated elevation and depression solitons. Thus we set these solitons to have random amplitudes varying in $[1, 3]$ and $[-1, -3]$ respectivaly.  The example of the initial soliton gas is presented in Figure \ref{Bipolardist}. 

\begin{figure}[h!]
\centering	
\includegraphics[scale =1]{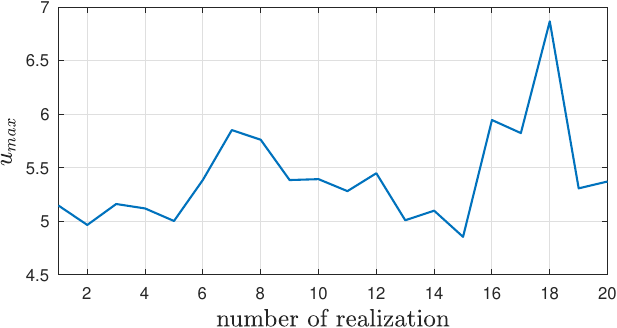}
\caption{Maximum of wave amplitudes over time of the bipolar soliton} 
\label{Maximumwave}
\end{figure}
\begin{figure}[h!]
\centering	
\includegraphics[scale =1]{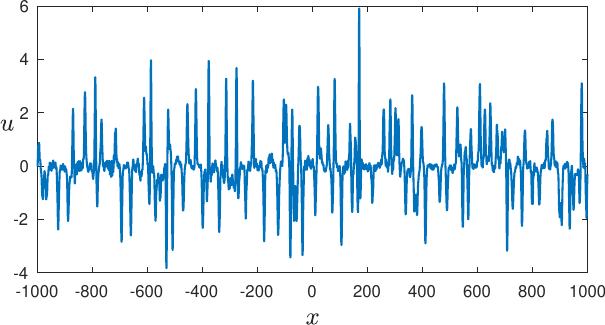}
\includegraphics[scale =1]{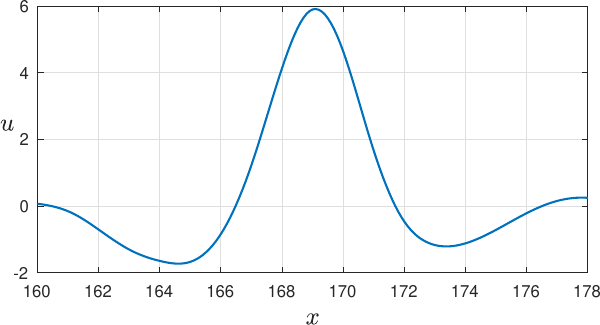}
\caption{Wave field of bipolar solitons at $t=416$. Zoom of the freak wave is on the bottom.} 
\label{freak}
\end{figure}
\begin{figure}[h!]
\centering	
\includegraphics[scale =1]{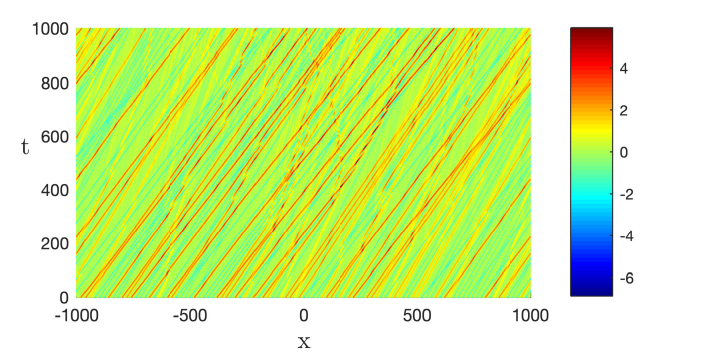}
\includegraphics[scale =1]{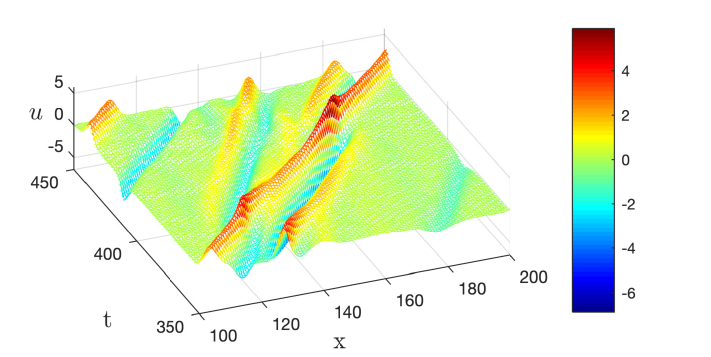}
\caption{Spatial-temporal diagram of bipolar soliton gas and zoom in the region of freak wave formation.} 
\label{meshfreak}
\end{figure}

Solitons interact with each other over time and there are collisions of big number of solitons. It may contribute to significant growth of the wave field. Thus, in considered realizations the maximum wave field reached almost 7.0 (Figure \ref{Maximumwave}). The wave field, which contais this freak wave is presented in Figure \ref{freak} and more details  of its formation is depicted in Figure \ref{meshfreak}.

\begin{figure}[h!]
\centering	
\includegraphics[scale =1]{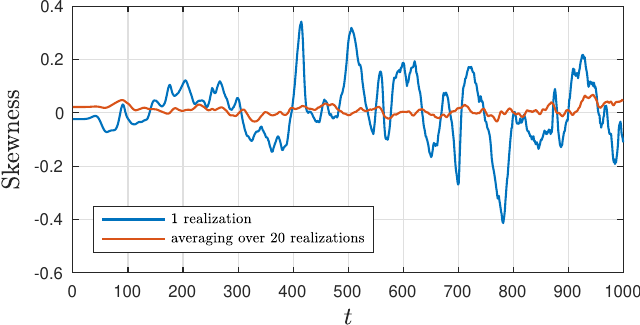}
\includegraphics[scale =1]{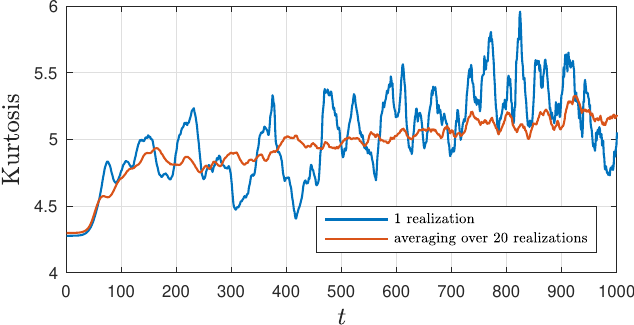}
\caption{Temporal evolution of the skewness and kurtosis of the bipolar soliton.} 
\label{kurtosisbi}
\end{figure}

\begin{figure}[h!]
\centering	
\includegraphics[scale =1]{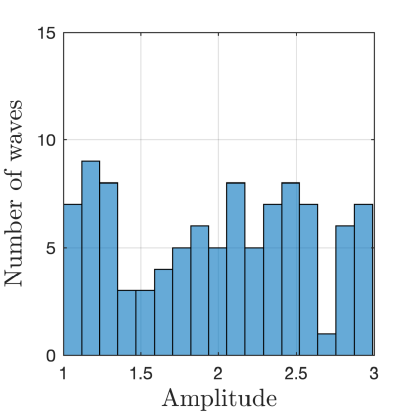}
\includegraphics[scale =1]{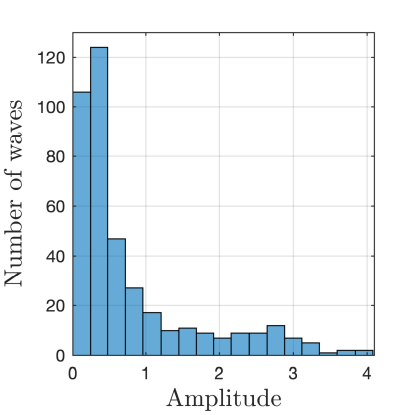}
\caption{Histogram of wave amplitude at $t=0$ (left) and at $t=1000$ (right).} 
\label{HistoBi}
\end{figure}

\begin{figure}[h!]
\centering	
\includegraphics[scale =1]{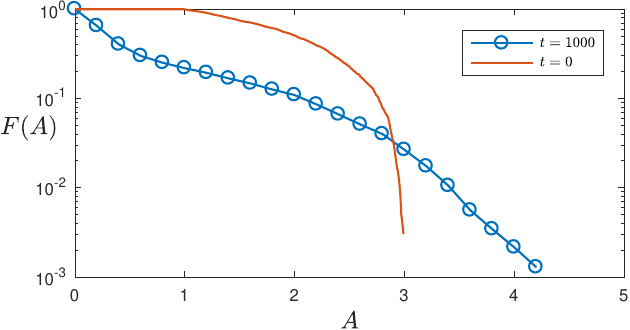}
\caption{Distribution function of wave amplitudes at different times averaged over 20 realizations.} 
\label{DistributionBi}
\end{figure}

In the process of nonlinear interaction skewness changes from $-0.4$ to $0.35$ in one realization, but the ensemble of bipolar solitons is close to be symmetrical, and the average value of this statistical moment takes values around zero. Skewness takes both positive and negative values depending on prevalence of positive and negative solitons. Kurtosis being the normalized integral of the wave field in fourth power takes values from $4.3$ to $6$ in one realization, however the deviation of the averaged value is much less. Moreover, Figure \ref{kurtosisbi} demonstrates the gradual growth of the averaged value of kurtosis. It may be explained by increasing of number of small amplitude waves (Figure \ref{HistoBi}), thus large waves become a ``more extreme" and kurtosis gradually increases. This behavior of kurtosis differs from the KdV model, where averaged kurtosis reached the steady state. The increase in kurtosis indicates the amplification of the distribution function tails (Figure \ref{DistributionBi}).

\section{Conclusion}
In this study, we have investigated  the dynamics of soliton gases using the Schamel equation as our primary framework. Our research involved a numerical exploration of various statistical characteristics, including distribution functions and moments, for both unipolar and bipolar soliton gases. Despite the non-integrability of the Schamel equation, the degree of dispersion generated during soliton interactions remained relatively modest, especially in the unipolar scenario. Consequently, our findings closely aligned with the predictions of the KdV and mKdV equations in the case of unipolar solitons. Nevertheless, in the scenario with bipolar characteristics, we noticed a significant deviation from the mKdV model, especially in terms of kurtosis behavior. The observed rise in kurtosis indicates the enhancement of distribution function tails, indicating the existence of high-amplitude waves.

\section{Acknowledgements}
M.V.F is grateful to IMPA for hosting him as visitor during the 2023 Post-Doctoral Summer Program. E.P. and E.D. are supported by Laboratory of Nonlinear Hydrophysics and Natural Disasters of the V.I. Il'ichev Pacific Oceanological Institute, grant from the Ministry of Science and Higher Education of the Russian Federation, agreement number 075-15-2022-1127 from 01.07.2022.

	\section*{Declarations}
	
	\subsection*{Conflict of interest}
	The authors state that there is no conflict of interest. 
	\subsection*{Data availability}
	
	Data sharing is not applicable to this article as all parameters used in the numerical experiments are informed in this paper.

\end{document}